\documentclass[twocolumn]{emulateapj} 
\usepackage{natbib}
\usepackage{epsfig}
\usepackage{graphicx} 
\usepackage{subfigure}
\usepackage{float}
\usepackage{amsmath}
\usepackage{color} 
\usepackage{amssymb}
\usepackage{amsfonts}
\usepackage{units}
\usepackage{mathtools}
\usepackage{bm} 
\usepackage[colorlinks,linkcolor=blue,anchorcolor=green,citecolor=blue]{hyperref} 
\def\be{\begin{eqnarray}}
\def\ee{\end{eqnarray}}

\shorttitle{FRBs as strong waves interacted with ambient medium}
\shortauthors{Yang \& Zhang}
\begin{document} 

\title{Fast radio bursts as strong waves interacting with ambient medium}

\author{Yuan-Pei Yang\altaffilmark{1}, and Bing Zhang\altaffilmark{2}}

\affil{
$^1$ South-Western Institute for Astronomy Research, Yunnan University, Kunming, Yunnan, P.R.China; ypyang@ynu.edu.cn;\\
$^2$ Department of Physics and Astronomy, University of Nevada, Las Vegas, NV 89154, USA; zhang@physics.unlv.edu
}

\begin{abstract}
Fast radio bursts (FRBs) are mysterious radio transients whose physical origin is still unknown. Within a few astronomical units near an FRB source, the electric field of the electromagnetic wave is so large that the electron oscillation velocity becomes relativistic, which makes the classical Thomson scattering theory and the linear plasma theory invalid. We discuss FRBs as strong waves interacting with the ambient medium, in terms of both electron motion properties and plasma properties. Several novel features are identified. 1. The cross section of Thomson scattering is significantly enhanced for the scattering photons. 2. On the other hand, because of the nonlinear plasma properties in strong waves, the near-source plasma is more transparent and has a smaller effective dispersion measure (DM) contribution to the observed value. For a repeating FRB source, the brighter bursts would have somewhat smaller DMs contributed by the near-source plasma. 3. The radiation beam undergoes relativistic self-focusing in a dense plasma, the degree of self-focusing (or squeezing) depends on the plasma density. Such a squeezing effect would affect the collimation angle and the true event rate of FRBs. 4. When an FRB propagates in a nearby ambient plasma, a wakefield wave in the plasma will be generated by the ponderomotive force of the FRB and accelerates electrons in the ambient medium. However, such an effect is too weak to be observationally interesting.
\end{abstract}

\keywords{radiation mechanisms: non-thermal}

\section{Introduction}\label{intro}

Fast radio bursts are mysterious radio transients with milliseconds-durations and large dispersion measures (DMs) \citep[e.g.][]{lor07,tho13,cha17,chime19a,chime19b,ban19,rav19}. The isotropic all-sky distribution and DM excess with respect to the Galactic contribution suggest that they are at cosmological distances \citep{tho13,sha18}. Precise localizations of a few FRBs firmly confirmed this \citep{cha17,ban19,rav19,pro19}. 
This implies very high luminosities of the bursts, which together with the short durations, imply  extremely high brightness temperatures of the order $T_{\rm B}\sim 10^{35}~{\rm K}$. Hence, the radiation mechanism of FRBs are required to be extremely coherent. Possible models include bunching curvature radiation \citep{kat14,kat18,kum17,ghi17,yan18} and maser mechanisms \citep{lyu14,bel17,bel19,ghi17b,lu18,met19}. The emission is so far only detected in a narrow band around $\sim1~{\rm GHz}$. The spectral extension of FRB emission to higher energies (both the same emission mechanism and the self-Compton emission) during the prompt emission phase (e.g. fast optical bursts) is likely weak \citep{yan19}, but it is possible that FRB sources may produce transients in other bands for different progenitor models \citep[see, e.g.][]{pla19}.

As radio transients at cosmological distances, FRBs can serve as tools for studying
the intergalactic medium (IGM), the interstellar medium (ISM), and the near-source plasma via the classical linear plasma theories \citep[e.g.][]{den14,rav16,xu16,yan17a,zha18,pro19}.
However, since the electric field of the electromagnetic wave is very large at the ambient medium of an FRB source, the oscillation velocity $v_{\rm os}$ of the accelerated electrons should be relativistic, i.e., $v_{\rm os}\sim c$, \citep{lua14,lyu17,bel19,lyu19,mar19,gru19,kum19,luph19}. In this case, the classical Thomson scattering theory and the linear plasma theory become invalid, and some peculiar properties, e.g., cross section enhancement, self-induced transparency, relativistic self-focusing, and wakefield acceleration, will play important roles in delineating the propagation properties of the strong waves \citep[e.g.][]{zel75,gib05,esa09,mac13}.

In this work, we discuss FRBs as strong waves interacting with the ambient medium and the corresponding observation properties. We first discuss the condition for FRB emission as strong waves in Section \ref{strongwave}. We then analyze the particle (Section \ref{electron}) and plasma (Section \ref{plasma}) properties in strong waves in Section \ref{pnp}. 
The results are summarized in Section \ref{discussion} with some discussion.

\section{Fast radio bursts as strong waves}\label{strongwave}

We consider an FRB source at a distance $d$ from Earth. Then its flux density at a distance $r$ from the source can be written as 
\be
F_\nu=\left(\frac{d}{r}\right)^2 S_\nu,
\ee
where $S_\nu$ is the observed FRB flux density. The corresponding electric field strength of the electromagnetic wave is 
\be
E\sim\left(\frac{4\pi\nu F_\nu}{c}\right)^{1/2}.
\ee
In the classical Thomson scattering theory, the motion of an electron is considered as non-relativistic, and is mainly affected by the electric field force in the wave. However, at high intensities, since electrons are accelerated to relativistic speeds, the Lorentz force is comparable to the electric force, e.g., $(e/c)(\bm{v}\times\bm{B})\sim eE$ for $v\sim c$. In this case, the electron motion is regulated by the electric and Lorentz forces together. Meanwhile, the corresponding motion becomes relativistic and is a nonlinear function of the driving field. 
In order to describe the strong-wave effect, one generally defines the strength parameter as
\be
a&=&\frac{v_{\rm os}}{c}=\frac{eE}{m_ec\omega}=\frac{eS_\nu^{1/2}d}{\pi^{1/2}m_ec^{3/2}\nu^{1/2}r}\nonumber\\
&=&1.2\left(\frac{S_\nu}{{\rm Jy}}\right)^{1/2}\left(\frac{\nu}{{\rm GHz}}\right)^{-1/2}\left(\frac{d}{{\rm Gpc}}\right)\left(\frac{r}{{\rm AU}}\right)^{-1}\label{strength},
\ee
where $v_{\rm os}= eE/m_e\omega$ is the typical oscillation velocity due to the electric force.
For a wave with $a\ll1$, the classical treatment of the Thomson theory is valid. However, for $a\gtrsim1$, one enters the regime of ``strong waves''. The relativistic motion of electrons and the Lorentz force contributed by the waves must be considered. According to Eq.(\ref{strength}), a critical radius $r_c$ is defined via $a(r_c)\equiv1$, which gives \citep{lua14}
\be
r_c&=&\frac{eS_\nu^{1/2}d}{\pi^{1/2}m_ec^{3/2}\nu^{1/2}}\nonumber\\
&\simeq&1.8\times10^{13}~{\rm cm}\left(\frac{S_\nu}{{\rm Jy}}\right)^{1/2}\left(\frac{\nu}{{\rm GHz}}\right)^{-1/2}\left(\frac{d}{{\rm Gpc}}\right)\label{radius}.
\ee
When an FRB propagates at $r\lesssim r_c\sim{\rm a~few~AU}$, its interactions with electrons microscopically and with plasma macroscopically must be treated with the strong-wave theory for $a\gg1$. We may call it the ``FRB-within-AU'' problem. 
Considering that the scale of the emission region of an FRB satisfies $r\gtrsim c\tau\simeq3\times10^7~{\rm cm}(\tau/1~{\rm ms})$, according to Eq.(\ref{strength}), the upper limit of the strength parameter satisfies 
\be 
a\lesssim6\times10^5\left(\frac{S_\nu}{{\rm Jy}}\right)^{1/2}\left(\frac{\nu}{{\rm GHz}}\right)^{-1/2}\left(\frac{d}{{\rm Gpc}}\right)\left(\frac{\tau}{1~{\rm ms}}\right)^{-1}.
\ee

\section{Particles and plasma in strong waves}\label{pnp}
\subsection{Free electron in strong waves}\label{electron}

In this section, we briefly discuss the electron motion properties in strong waves. The strong-wave problem has been solved in the classical theory with exact allowance for relativistic mechanics and Lorentz forces \citep{sar70}. We assume that an electron is initially at rest at the origin in the laboratory, and the incident wave is transverse, plane, and elliptically polarized.
As the electromagnetic waves pass across an electron, the electron would move due to the electromagnetic interaction, and its motion contains three components: 
\begin{itemize}
\item the classical harmonic motion transverse to the wave direction due to the electric field; 
\item the longitudinal harmonic motion due to the Lorentz force;
\item the drift motion along the propagation direction of the waves\footnote{For a plane wave with an infinite duration, the oscillation center of the electron is always at rest according to the Thomson scattering theory. Realistically, for an electromagnetic pulse with a finite duration, a ponderomotive force from the electromagnetic pulse would accelerate the oscillation center, leading to a drift velocity along the propagation direction of the waves.}. 
\end{itemize}
After a wave pulse passes by, the harmonic motion and the drift velocity die out due to the ponderomotive force. The electron  again becomes at rest in the laboratory. For strong waves with a frequency $\omega$ and a duration $T\gg1/\omega$, the drift velocity along the direction of the waves is \citep{sar70}
\be
v_D\simeq \frac{a^2}{a^2+4}c,
\ee
which corresponds to a Lorentz factor of $\Gamma_D\simeq a/2\sqrt{2}$ for $a\gg1$. 
In the oscillation-center rest frame (where the cycle-averaged position is at rest, i.e., $v_D=0$), the motion of the electron is relativistic with a mean Lorentz factor of $\bar\gamma'\sim a$ for $a\gg1$.

\begin{figure}[]
\centering
\includegraphics[angle=0,scale=0.25]{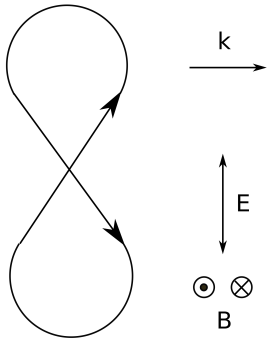}
\caption{The ``figure-of-eight'' trajectory of an electron under the strong linearly polarized wave. The arrows on the side denote the polarization of the wave.}\label{fig1} 
\end{figure}

Let us define that the incident wave travels along the $+z$ direction, that the classical harmonic oscillation by the electric field force is in the $xy$ plane, that the second harmonic oscillations by Lorentz force is in the $z$ direction, and that the drift velocity $v_D$ is along the $+z$ direction. 
In particular, for circular polarization, the oscillating $z$ motion vanishes so that the resulting orbit is helical. In the oscillation-center rest frame, the electron orbit is circular in the $xy$ plane with a constant local Lorentz factor $\gamma'=a/\sqrt{2}$ \citep{sar70}.
For linearly polarization, the orbit is a ``figure-of-eight trajectory'' in the $xz$ plane in the oscillation-center rest frame, and the electron moves slowest on the round part of the orbit and fastest on the straight part \citep{sar70}, as shown in Figure \ref{fig1}.  
Such a figure-of-eight trajectory has been confirmed in the experiments of the nonlinear Thomson scattering \citep{che98}.
As the parameter $a$ gets smaller, the orbit approaches the one-dimensional harmonic oscillator in the low-intensity treatment of Thomson scattering.

We should note that the above discussion assumes that the electron motion is in an external electromagnetic field that is independent of the electron motion. However, the electromagnetic field radiated by the electron itself would react on the electron dynamics, which is known as ``radiation friction''. Due to the radiation friction force, in the oscillation-center rest frame the electron trajectory would open up \citep[e.g.][]{zel75,mac13}.

The electron motion determines its radiation. In the oscillation-center rest frame, the electron motion is relativistic with the mean Lorentz factor of $\bar{\gamma}'\sim a$ for $a\gg1$. 
The radiation is confined to a narrow cone of $1/\gamma'$ along the electron velocity, and the corresponding spectrum therefore contains higher harmonics of the fundamental frequency with which the electron rotates. Similar to synchrotron radiation, the maximum harmonic number would be $m\sim\gamma'^3\sim a^3$, which corresponds to a critical frequency of $\omega_c'\sim a^3\omega_0'$, where $\omega_0'$ is the fundamental frequency in the rest frame. In the observer frame, due to the relativistic drift motion with Lorentz factor of $\Gamma_D\sim a$, according to the Doppler effect, the observed frequency would become $\omega\sim a\omega'$ for $a\gg1$ along the line of sight. Meanwhile, the relativistic drift motion would make radiation predominantly forward and confined to an angle of $\theta\sim1/\Gamma_D\sim1/a$.
At last for $a\gg1$, the mean total received power by an observer is larger than that given by the Thomson formulas by approximately a factor $a^2$ \citep{sar70}, i.e., 
\be
P\sim a^2P_T,\label{power}
\ee 
where $P_T=e^4E^2/3m_e^2c^3$ is the mean received power given by the Thomson formula. Notice that in Eq.(\ref{power}) $P$ is the received power that has been corrected by the retardation effect \citep[e.g.][]{ryb79}. If one focuses on the emitted power $P_e$, one has $P_e\sim P_T$ as pointed out by \citet{gun71}.
Considering that the energy flux in the waves is $S=cE^2/8\pi$, the cross section for scattering photons is approximately given by 
\be
\sigma=\frac{P}{S}\sim a^2\sigma_T\simeq\frac{8\pi e^6E^2}{3m_e^4c^6\omega^2}
\ee
Therefore, for $a\gg1$, the cross section for scattering photons would be much larger than the Thomson cross section by a factor of $\sim a^2$. After scattering, the forward radiation of the electron in the propagation direction of the strong waves would weaken the injection waves because of the energies scattered to other directions. 

At last, we discuss the ponderomotive force contributed by a pulse of electromagnetic waves. For an infinite monochromatic plane wave, the oscillation center of an electron is always at rest.
The ``realistic'' electromagnetic fields are not perfectly monochromatic plane waves, but have finite widths and durations.
For a wave with a finite width and duration, besides the fast harmonic motion, the oscillation center would be accelerated by the ponderomotive force. 
In the non-relativistic regime, the ponderomotive force is 
\be
\bm{F}_p=-\nabla\Phi=-\frac{1}{2}m_ec^2\nabla\left<\bm{a}^2\right>\label{pond}
\ee
where $\Phi=e^2\left<\bm{E}\right>/2m_e\omega^2$ is the ponderomotive potential, $\bm{a}=e\bm{A}/m_ec^2$, and $\bm{A}=\nabla\times\bm{B}$ is the vector potential. The result of ponderomotive force is that electrons will be expelled from the regions where the electric field is higher, which can be viewed as the radiation pressure.
In the relativistic regime, the ponderomotive force is \citep{bau95,mul10}
\be
\bm{F}_p=-m_ec^2\nabla\left(1+\left<\bm{a}^2\right>\right)^{1/2}.
\ee
Similar to the non-relativistic regime, for the relativistic regime ($a\gtrsim1$) electrons would be scattered off from regions where the electric field is higher. Due to the ponderomotive force, a wakefield in plasma would form and electrons in plasma would be accelerated by the wakefield wave, as discussed in Section \ref{wakefield}.

\subsection{Plasma in strong waves}\label{plasma}

In this section, we consider the plasma properties under the propagation of strong waves. In strong waves, the motion of electrons in the plasma becomes relativistic. 
However, different from free electrons that have a relativistic drift velocity in the direction of the incident electromagnetic wave (see Section \ref{electron}), in plasma the space-charge potential is important in preventing the drift of electrons \citep{wal78}. For non-relativistic electrons in plasma, if the wave duration $\tau$ is much larger than $c/\omega_p$, where $\omega_p=\sqrt{4\pi e^2n_e/m_e}$ is the plasma frequency, the drift velocity would be close to zero \citep{wal78,spr90}. In this case, electrons in plasma under a strong wave would have a typical Lorentz factor ($\gamma'$) similar to that ($\gamma$) in the laboratory frame, so that $\gamma\sim\gamma'\sim a$ is satisfied.
Due to the relativistic and magnetic force effects, the propagation and dispersion properties of an electromagnetic wave depend on its amplitude.
For a circular polarized wave, the disperse relation in the laboratory frame is given by \citep[e.g.,][see Appendix A]{gib05,mac13,mac15}
\be
\omega^2=k^2c^2+\frac{\omega_p^2}{\gamma},~~~\gamma=(1+a^2/2)^{1/2}.
\label{dispersion}
\ee
The dispersion relation of strong electromagnetic waves is altered due to the effective electron mass increased by the relativistic effect \citep[e.g.][]{sar70,gib05,mac13}. One can define the effective plasma frequency as
\be
\omega_{p,{\rm eff}}=\frac{\omega_p}{\sqrt{\gamma}},
\ee
so that the wave can propagate in the region where $\omega>\omega_{p,{\rm eff}}=\gamma^{-1/2}\omega_p$. With respect to the non-relativistic linear case, this is known as relativistically self-induced transparency.
We note that since the dispersion depends on the electromagnetic field amplitude in the nonlinear case, the dispersion relation must be taken with care. The propagation of a pulse will be affected by the complicated effects of nonlinear propagation and dispersion, and finally the spatial and temporal shape of the pulse itself would also be modified. In particular, for linear polarization, the relativistic factor $\gamma$ is not a constant (see Section \ref{electron}). The propagation of the linearly polarized wave with a relativistic amplitude would lead to generation of the higher order harmonics. \citet{spr90} proved that the propagation of the first harmonic component, i.e. of the “main” wave, is still reasonably described by Eq.(\ref{dispersion}) with $\gamma\rightarrow\left<\gamma\right>$. Thus, we will directly adopt Eq.(\ref{dispersion}) in the following discussion.

According to Eq.(\ref{dispersion}), the effective cut-off electron density is 
\be
n_c=\frac{\gamma m_e\omega^2}{4\pi e^2}\simeq \frac{m_e\omega^2 a}{2^{5/2}\pi e^2},~~~\text{for}~~~a\gg1.
\ee
The plasma with electron number density $n_e<n_c(\omega)$ would be transparent for the electromagnetic waves with a frequency $\omega$. 
According to the dispersion relation given by Eq.(\ref{dispersion}), the group velocity of the electromagnetic waves is
\be
v_g=\frac{\partial\omega}{\partial k}=c\left[1-\frac{\omega_p^2}{\omega^2(1+a^2/2)^{1/2}}\right]^{1/2}
\simeq c\left[1-\frac{1}{\sqrt 2a}\frac{\omega_p^2}{\omega^2}\right], \nonumber\\\label{gvel}
\ee
where $\omega\gg\omega_p$ and $a\gg1$ is assumed.
Replacing $\omega_p$ with $\omega_{p,{\rm eff}}$, for an ambient medium with a free electron number density $n_e(r)$, the effective DM with $r< r_c$ is
\be
{\rm DM_{eff}}=\int_{r_e}^{r_c}\frac{n_e(r)}{\gamma} dr\simeq\frac{\sqrt 2}{a_0}\int_{r_e}^{r_c}n_e(r)\frac{r}{r_e}dr,
\ee
where $r_e$ is the radius of the emission region, $r_c$ is the critical radius given by Eq.(\ref{radius}), and the strength parameter is assumed to be $a=a_0(r_e/r)$ due to the inverse-square law of the flux. For the uniform ambient medium, the effective DM is ${\rm DM_{eff}}=n_er_c^2/\sqrt 2 a_0r_e\sim n_e r_c/\sqrt 2\sim {\rm DM}$ due to $a_0 r_e/r_c\sim1$. Thus, the effective DM with $r_c$ is of the order of magnitude of the classical DM, because most DM is contributed by the plasma at the scale of $\sim r_c$ (corresponding to $a\sim1$). However, for the wind medium with $n_e(r)=n_{e,0}(r_e/r)^2$, the effective DM, i.e., ${\rm DM_{eff}}=(\sqrt 2/a_0)n_{e,0}r_e\ln(r_c/r_e)$, would be much smaller than the classical DM with ${\rm DM}=\int n_e(r)dr\simeq n_{e,0}r_e$ for $a_0\gg1$. The DM contribution near the FRB source would be significantly suppressed due to the strong-wave effect. This is relevant to, for example, synchrotron maser models invoking relativistic magnetized shocks in a steady magnetar wind with $n_e\propto r^{-2}$ \citep[e.g.][]{bel17,bel19,met19,mar19}.
In this case, the dispersion relation would involve the strong-wave effect, leading to the near-source plasma more transparent and therefore a smaller effective DM. Some FRBs, e.g, FRB 180924 \citep{ban19}, have a small observed DM which might be mostly accounted for by the contribution from the intergalactic medium, implying a negligibly small DM from the FRB near-source plasma. Our results show that it is still possible that the near-source plasma column density is not too small as long as it is confined within 1 AU from the source with a stratified wind density profile. 
At last, as discussed above, for a repeating FRB source, considering the strong-wave effect, the brighter bursts would have somewhat smaller DMs contributed by the near-source plasma\footnote{\cite{luph19} independently reached the similar qualitative conclusion even though quantitatively they used a different dispersion relation.}. In particular, the DMs contributed by the near-source plasma with a few times of $r_e$ satisfies ${\rm DM}\propto a_0^{-1}\propto \nu^{1/2}S_{\nu}^{-1/2}$, where $\nu$ is the burst frequency, and $S_\nu$ is the observed burst flux density.

\subsubsection{Relativistic Self-Focusing}\label{selffocus}

\begin{figure}[H]
\centering
\includegraphics[angle=0,scale=0.3]{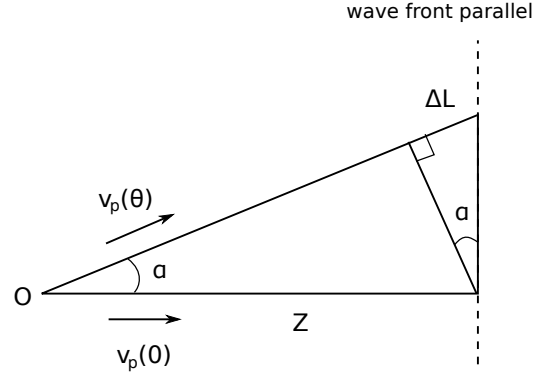}
\caption{Geometry for self-focusing effect. The electromagnetic waves are emitted at point $O$, and the intensity at the beam center is larger than that at $\theta$, i.e., $a(0)>a(\theta)$. According to Eq.(\ref{phase}), the phase velocity at the beam center, $v_p(0)$, will be smaller than that at $\theta$, $v_p(\theta)$. After propagating a distance of $Z$, the global wave becomes planar. $\alpha$ corresponds to the focusing angle, and $\Delta L$ corresponds to the light path difference.}\label{fig2}
\end{figure}

The dispersion relation, Eq.(\ref{dispersion}), predicts that a strong wave would undergo self-focusing for a structured beam.
According to Eq.(\ref{dispersion}), the phase velocity is given by
\be
v_p=\frac{\omega}{k}=c\left[1-\frac{\omega_p^2}{\omega^2(1+a^2/2)^{1/2}}\right]^{-1/2}
\simeq c\left[1+\frac{1}{\sqrt 2a}\frac{\omega_p^2}{\omega^2}\right]\nonumber\\\label{phase}
\ee
where $\omega\gg\omega_p$ and $a\gg1$ is assumed. We consider that the FRB radiation is beamed with a typical angle of $\theta_0$: the flux is maximum at the center of the beam and decreases towards edge, i.e., $a(\theta_0)< a(0)$. The nonlinear refractive index is $n=c/v_p=\sqrt{1-\omega_p^2/\gamma(a)\omega^2}$, which is intensity-dependent. This suggests that the refractive index is maximum at $\theta=0$ and decreases with $\theta$.
The phase velocity difference $\Delta v_p$ between wave A at $\theta=0$ and wave B at $\theta=\theta_0$ is
\be
\frac{\Delta v_p}{c}=\left(\frac{1}{a(\theta_0)}-\frac{1}{a(0)}\right)\frac{\omega_p^2}{\sqrt2\omega^2}\simeq\frac{\omega_p^2}{\sqrt2\omega^2}\frac{1}{a(\theta_0)}. \nonumber\\
\ee
Here $a(\theta_0)\ll a(0)$ is assumed. As shown in Figure \ref{fig2}, the maximum light path difference between $\theta=0$ and $\theta=\theta_0$ is
\be
\Delta L\simeq\left|\Delta v_p/c\right|_{\max}Z\simeq\alpha^2Z.
\ee
As shown in Fig.\ref{fig2}, the phase velocity at the angle $\alpha$ is greater than that in the zero angle direction, so that at distance $Z$ the global wave becomes planar, with the wavefront denoted as the dashed line.
Therefore, the focusing angle of the beam is given by
\be
\alpha\simeq\left|\frac{\Delta v_p}{c}\right|_{\max}^{1/2}\simeq\frac{1}{2^{1/4}a^{1/2}(\theta_0)}\frac{\omega_p}{\omega}
\ee
for $r\lesssim r_c$.
The maximum focusing angle corresponds to\footnote{For weak waves with $\omega\gg\omega_p$ and $a(\theta)\ll1$ at $0<\theta<\theta_0$, similar to the above discussion, the focusing angle is $\alpha\simeq(a(0)/2\sqrt2)(\omega_p/\omega)$.} $a(\theta_0)\sim1$, thus one has
\be
\alpha_{\max}\simeq\frac{1}{2^{1/4}}\frac{\omega_p}{\omega}\simeq0.8\left(\frac{\nu}{1~{\rm GHz}}\right)^{-1}\left(\frac{n_e}{10^{10}~{\rm cm^{-3}}}\right)^{1/2}
\ee

We consider that a radiation beam has an original beaming angle of $\theta_j$. Because of the relativistic self-focusing effect, the beaming angle would squeeze and become $\sim(\theta_j-\alpha)$. 
In some FRB models invoking synchrotron maser emission from relativistic blast waves \citep[e.g.][]{met19,bel19,mar19}, the electron number density near the maser emission region could be high for the wind external medium. We assume that the intrinsic beaming angle is $\theta_j$. For $\theta_j\sim\alpha$, i.e., $n_e\sim2\times10^8~{\rm cm^{-3}}(\theta_j/0.1)^2(\nu/1~{\rm GHz})^2$, the FRB beam would be squeezed by the relativistic self-focusing effect significantly, leading to a smaller observation probability than the classical picture. If most FRBs are affected by the squeezing effect, the true event rate density would become $\rho\propto(\theta_j-\alpha)^{-2}$, in which case the true event rate density would be much larger than that constrained by the current observations \citep[e.g.][]{cao18}. Furthermore, if $\theta_j<\alpha$, the FRB propagation would be similar to what happens in an optical fiber. Such an FRB is almost impossible to detect due to the extremely narrow beam.

\subsubsection{Electron acceleration in wakefield waves}\label{wakefield}

In general, the dispersion relation of electrostatic waves in a cold plasma is $\omega=\omega_p$ for any wave-vector $k$. Thus, the wavelength of the electrostatic waves in plasma is determined by the way the wave is excited. We consider that a charged particle is accelerated by the ponderomotive force traveling in the plasma at a velocity $v_f$. As discussed in Section \ref{electron}, the ponderomotive force would cause the charged particle expelled from the regions where the electric field is higher, similar to the effect of radiation pressure, i.e., $\bm{F}_p\propto -m\nabla\left<\bm{a}^2\right>$ for $a\ll1$ and $\bm{F}_p\propto -m\nabla\left<\bm{a}\right>$ for $a\gtrsim1$, where $m$ is the particle mass. Since the electron mass is much smaller than the proton mass, electrons are easier to be accelerated than protons. When the electrons are away from the equilibrium positions, an electrostatic field is generated in the plasma, leading to the generation of an electrostatic wave due to plasma oscillation. This is the so called wakefield waves. Such an effect was first proposed by \citet{taj79} and has been expensively applied to the field of laser-driven plasma acceleration \citep[reviewed by][]{esa09}.
Since the oscillation is produced at the ponderomotive force front, the phase velocity of the wakefield wave is equal by construction to the velocity of the force perturbation, i.e., $v_p=\omega_p/k=v_f$. 

\begin{figure}[H]
\centering
\includegraphics[angle=0,scale=0.35]{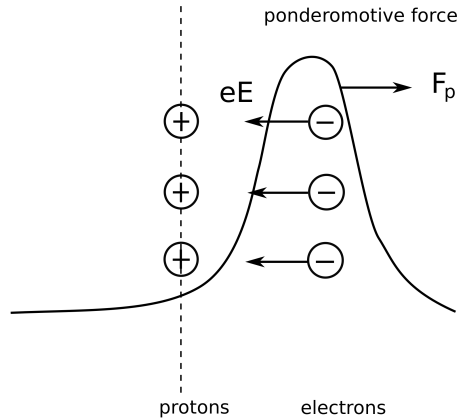}
\caption{Wakefield excitation. The curvature line represents the pulse of an electromagnetic wave. The positive circles represents protons, and the negative circles represents electrons. $eE$ is the electric force, and $F_p$ is the ponderomotive force by the electromagnetic wave pulse. When the electromagnetic wave pulse propagates in the plasma, the ponderomotive force from the wave pulse would accelerate the electrons in the plasma. At the pulse front, the ponderomotive force points to the direction of the pulse propagation. At the pulse tail, the ponderomotive force points to the anti-direction of the pulse propagation. After the electromagnetic wave pulse propagates in plasma, an electrostatic wave would be generated due to the plasma oscillation, which is called the wakefield wave.}\label{fig3}
\end{figure}

We first consider electron acceleration in the wakefield wave of the incident electromagnetic wave with $a\ll1$. In the plasma, the total electron number density is positive, i.e., $n_e=n_{e,0}+\delta n_e\geqslant0$, leading to $|\delta n_e|\leqslant n_{e,0}$. 
Assuming that the wakefield wave propagates along the $x$ axis, according to the Gauss' law, e.g, $\partial_x E_x=-4\pi e\delta n_e$, using $\partial_x\sim k=\omega_p/v_p$ where $\omega_p=(4\pi e^2 n_{e,0}/m_e)^{1/2}$ is the plasma frequency and $v_p$ is the phase velocity of the wakefield wave, one has the electric field strength of the wakefield wave satisfying
$|E_x|\leqslant E_0=m_ev_p\omega_p/e$. 
In an underdense plasma, the group velocity of the electromagnetic wave is $v_g^{\rm EM}\simeq c$. Meanwhile, the wakefield wave is excited by the ponderomotive force created by the electromagnetic wave.
Thus, the wakefield wave has the phase velocity of $v_p=v_g^{\rm EM}\simeq c$ for $\omega\gg\omega_p$. The upper limit of the wakefield wave is determined by
\be
E_0\simeq\frac{m_ec\omega_p}{e}\label{E0}.
\ee
According to the properties of the ponderomotive force, i.e., $\bm{F}_p\propto -m\nabla\left<\bm{a}^2\right>$ for $a\ll1$, an electron is accelerated at the wavefront, and decelerated at waveback, see Figure \ref{fig3}. When the acceleration timescale of the ponderomotive force is approximately equal to the plasma oscillation timescale, the resonance condition would be satisfied.
Thus, the wakefield wave is most effectively generated when the electromagnetic wave pulse length $\sim c\tau$ is roughly matched to the wavelength $\lambda_p$ of the wakefield wave \citep{taj79}, i.e.,
\be
c\tau\sim\lambda_p=\frac{2\pi c}{\omega_p}
\ee
In particular, for an FRB with a duration of $\tau\simeq1~{\rm ms}$, the typical electron number density for the enhanced wakefield wave is approximately
\be
n_e\simeq\frac{\pi m_e}{ e^2\tau^2}\simeq0.01~{\rm cm^{-3}}\left(\frac{\tau}{1~{\rm ms}}\right)^{-2}. \label{ismdens}
\ee
It is interesting that such an electron number density is close to the typical number density of the interstellar medium (ISM). Therefore, when an FRB propagates in the ISM, a wakefield wave would be effectively generated, especially in the region with a density $n_e\sim0.01~{\rm cm^{-3}}$.
We notice that Eq.(\ref{E0}) gives the maximal electric field in the wakefield wave, which implies a complete charge separation. As the electromagnetic wave propagates, the ponderomotive force becomes weak, leading to an incomplete charge separation and a weaker maximum electric field in the wakefield wave satisfying $eE_{\rm max}\sim F_p$. According to Eq.(\ref{pond}), the ponderomotive force is $F_p\sim m_eca^2/\tau_p\sim m_ec\omega_pa^2$ for $a\ll1$, where $\tau_p\sim 1/\omega_p$ is the typical timescale of the ponderomotive force acting on the plasma. According to $E_{\rm max}\sim F_p/e$ and Eq.(\ref{E0}), for $a\ll1$, the maximum electric field in the wakefield wave is 
\be
E_{\rm max}\sim a^2E_0.
\ee

The electric field in the wakefield wave would accelerate electrons in the plasma to relativistic velocities, leading to electron-trapping in the electrostatic waves when $v\simeq v_p\sim c$. In the wave frame, the relativistic electrons would be accelerated over at most half a wavelength in the wave-frame, after which it starts to be decelerated. Thus, the acceleration length is
\be
l_{\rm acc}\simeq\frac{\lambda_pc}{2|c-v_p|}\simeq\gamma_p^2\lambda_p=\lambda\left(\frac{\omega}{\omega_p}\right)^3,\label{lacc0}
\ee
where $\gamma_p=1/\sqrt{1-v_p^2/c^2}=\omega/\omega_p$ is the Lorentz factor of the wakefield wave related to the observer frame, and $v_p\simeq v_g^{\rm EM}=c\sqrt{1-\omega_p^2/\omega^2}$.
The above equation gives the maximum acceleration length in the uniform plasma. 

For $a\ll1$, the wakefield wave is a simple sinusoidal oscillation with a wavelength $\lambda_p$. However, in the ultrarelativistic limit in which the amplitude of the electromagnetic wave pulse satisfies $a\gg1$, the wakefield wave would be nonlinear, allowing $E_{\max}>E_0$ and an $a$-dependent wavelength $\lambda_p$  \citep{spr90b,spr90}. For a square electromagnetic pulse profile with $a\gg1$, the maximum electric filed of the wakefield wave is 
\be
E_{\max}\sim\frac{a^2}{\sqrt{1+a^2}}E_0\sim aE_0,
\ee
and the wavelength of the wakefield wave is 
\be
\lambda_{Np}\sim\lambda_p\left(\frac{E_{\max}}{E_0}\right)\sim a\lambda_p
\ee
for $E_{\max}\gg E_0$, where $\lambda_p=2\pi c/\omega_p$ \citep{spr90b,spr90,esa09}. The amplitude of the longitudinal oscillation would be enhanced if the pulse length is roughly matched to the wavelength of the wakefield wave, i.e.
\be
c\tau\sim\lambda_{Np}\sim\frac{2\pi c a}{\omega_p}.
\ee
For an FRB with a duration $\tau\simeq1~{\rm ms}$, the typical electron number density for an enhanced wakefield wave is
\be
n_e\simeq\frac{\pi m_ea^2}{ e^2\tau^2}\simeq0.01~{\rm cm^{-3}}a^2\left(\frac{\tau}{1~{\rm ms}}\right)^{-2}. \label{densa}
\ee
In order to make relativistic electrons accelerated over at most half a wavelength in the wave-frame, the acceleration length is required to be
\be
l_{\rm acc}\simeq\gamma_p^2\lambda_{Np}=a^2\lambda\left(\frac{\omega}{\omega_p}\right)^3\label{lacc1}
\ee
where $\gamma_p=1/\sqrt{1-v_p^2/c^2}\simeq a^{1/2}\omega/\omega_p$ is the Lorentz factor of the wakefield wave in the observer frame, and $v_p\simeq v_g^{\rm EM}=c\sqrt{1-\omega_p^2/\gamma\omega^2}$ according to Eq.(\ref{gvel}). 

In the above discussion, a uniform and large-scale plasma without magnetic field is assumed. 
However, in the real ISM with inhomogeneous gas density and magnetic field, the above maximum acceleration length is likely significantly suppressed. The reason is as follows: (i) In the ISM, the coherent length $l_{\rm coh}$ corresponding to the critical electron number density with $n_e$ given by Eq.(\ref{ismdens}) or Eq.(\ref{densa}) is finite. For $l_{\rm coh}\ll l_{\rm acc}$, the maximum acceleration length becomes $\sim l_{\rm coh}$. (ii) Due to the magnetic field in the ISM, the electrons would be accelerated by the electric field along the magnetic field line, if the Larmor radius of electrons satisfies $r_L \ll l_{\rm acc}$. The former is $r_L=\gamma m_e c^2/eB = 1.7\times 10^9 \ {\rm cm} \ \gamma (B/1~{\rm \mu G})^{-1}$. One can see the condition $r_L \ll l_{\rm acc}$ is readily satisfied.  Assume that the angle between the electric field and the magnetic field is $\theta$. For the ISM with a random magnetic field, the electrons will be accelerated along the magnetic field line, and the acceleration length becomes
\be
l_{\rm acc,B}\simeq\frac{\lambda_{p,a}c}{2|c\left<\cos\theta\right>-v_p|}.
\ee
where $\lambda_{p,a}\equiv\lambda_p$ for $a\ll1$ and $\lambda_{p,a}\equiv\lambda_{Np}$ for $a\gg1$.
Since $\left<\cos\theta\right>\simeq 1/2$ for a random field and $v_p\simeq c$, one has $l_{\rm acc,B}\sim\lambda_p$ for $a\ll1$ and $l_{\rm acc,B}\sim\lambda_{Np}\sim a\lambda_p$ for $a\gg1$. Thus $l_{\rm acc,B}$ is much less than $l_{\rm acc}$ given by Eq.(\ref{lacc0}) and Eq.(\ref{lacc1}). 
For both $a\ll1$ and $a\gg1$, the electrons could be accelerated to $\gamma\sim eE_{\max}l_{\rm acc,B}/m_ec^2\sim2\pi a^2$ by the wakefield wave, but the corresponding synchrotron radiation from accelerated electrons is extremely low. Therefore, the acceleration from the wakefield wave would become inefficient due to the external magnetic field. Such an effect is too weak to be observationally interesting

\section{Conclusions and Discussion }\label{discussion}

The strong-wave problem has been solved in the classical theory for a point charge \citep{sar70} and in the quantum theory \citep{bro64}. In strong waves, electrons would be accelerated to relativistic velocities, leading to modifications of the classical plasma properties  \citep[e.g.][]{spr90b,spr90}, including self-induced transparency, relativistic self-focusing and wakefield acceleration. These effects have been extensively applied to the laser-driven plasma acceleration experiments, as reviewed by \citep{gib05,esa09,mac13,mac15}. Similar to  laser propagation in plasma, the strong-wave effect can also play a significant role in the astrophysical processes, especially when strong radiation propagating in a dense near-source plasma. 
In this work, we discuss FRBs as strong waves interacting with the ambient medium. When an FRB propagates at $r\lesssim{\rm a~few}~{\rm AU}$ near the source, the electric field of the electromagnetic waves is so large that the electron oscillation velocity becomes relativistic, which makes the classical Thomson scattering theory and the the linear plasma theory invalid. 

For a free electron under strong waves with $a\gg1$, its motion would significantly deviate from the harmonic motion in the classical Thomson scattering theory, because the Lorentz force is almost equivalent to the electric force, i.e. $(e/c)(\bm{v}\times\bm{B})\sim e\bm{E}$ for $v\sim c$. In this case, the electron radiation power would be larger than that given by the Thomson formula by a factor of $a^2$, i.e. $\sigma\sim a^2\sigma_{\rm T}$, where $\sigma_{\rm T}$ is the Thomson scattering cross section. 

The plasma properties in strong waves are also discussed, including self-induced transparency, relativistic self-focusing and electron acceleration in the wakefield wave. Due to the nonlinear plasma properties, the effect plasma frequency becomes $\omega_{p,{\rm eff}}\sim a^{-1/2}\omega_p$. 
Thus, near the FRB source, the plasma would be more transparent than the  results predicted by the classical theory, and the corresponding effective DM becomes smaller. 
In particular, for a repeating FRB source, the brighter bursts would have somewhat smaller DMs contributed by the near-source plasma. The DMs from a few times of the emission radii satisfies ${\rm DM}\propto \nu^{1/2}S_{\nu}^{-1/2}$.
On the other hand, the nonlinear properties also cause an intensity-dependent refractive index. For an FRB with a structured beam, e.g., a decreasing intensity as an angle from the beam axis, the radiation beam would be relativistically self-focused in the near-source plasma. For an electron number density $n_e\gtrsim2\times10^8~{\rm cm^{-3}}(\theta_j/0.1)^2(\nu/1~{\rm GHz})^2$, where $\theta_j$ is the intrinsic FRB beaming angle and $\nu$ is the FRB frequency, the FRB beam would be squeezed by the self-focusing effect. The above effects might be important in some FRB models, such as the maser emission model in a relativistic outflow \citep{bel17,bel19,met19}, cosmic combs \citep{zha17b,zha18b}, FRB generation and propagation in a pulsar magnetosphere \citep{dai17,lu18,yan18,wan19,wan19b,yyh19}, etc. In these cases, the near-source plasma could be dense close to the FRB emission region. If most FRBs are affected by the squeezing effect, the true event rate density would become higher than that constrained by the current observations.

\acknowledgments 
We thank Pawan Kumar and Wenbin Lu for valuable comments and discussions and an anonymous referee for constructive criticisms.

\appendix

\section{Appendix A: Dispersion relation of strong waves in plasma}

In this section, we derive the dispersion relation of strong waves in plasma. First, we consider an electromagnetic wave propagating along $\hat{\bm{x}}$ with the vector potential $\bm{A}(x,t)$. By noticing $\bm{A}=\bm{A}_\perp$, the electron momentum in the transverse $yz$ plane is given by
\be
\frac{d\bm{p}_\perp}{dt}=e\bm{E}+\frac{e}{c}(\bm{v}\times\bm{B})_\perp=\frac{e}{c}\left(\frac{\partial\bm{A}}{\partial t}+v_x\frac{\partial\bm{A}}{\partial x}\right)=\frac{e}{c}\frac{d\bm{A}}{dt},
\ee
leading to
\be
\frac{d}{dt}\left(\bm{p}_\perp-\frac{e}{c}\bm{A}\right)=0.
\ee
Taking $\bm{p}_\perp=0$ and $\bm{A}=0$ at the initial time (assuming that the turn-on time is arbitrarily long, e.g., adiabatic rising), one finally has
\be
\bm{p}_\perp=\gamma m_e\bm{v}_\perp=\frac{e}{c}\bm{A}\label{momp},
\ee
where $\gamma$ is the electron Lorentz factor in the observer frame.
On the other hand, according to Maxwell’s equations, the electromagnetic wave equation satisfies
\be
\nabla^2\bm{A}-\frac{1}{c^2}\frac{\partial^2\bm{A}}{\partial t^2}=-\frac{4\pi}{c}\bm{J}+\nabla(\nabla\cdot\bm{A})+\frac{1}{c}\nabla\frac{\partial\phi}{\partial t}.
\ee
Split the current $\bm{J}$ into $\bm{J}_\parallel$ and $\bm{J}_\perp$, i.e., $\bm{J}=\bm{J}_\parallel+\bm{J}_\perp$. Applying the Coulomb gauge $\nabla\cdot\bm{A}=0$ and $\bm{J}_\parallel=(1/4\pi)\nabla(\partial\phi/\partial t)$, one finally has
\be
\nabla^2\bm{A}-\frac{1}{c^2}\frac{\partial^2\bm{A}}{\partial t^2}=-\frac{4\pi}{c}\bm{J}_\perp. 
\ee
According to Eq.(\ref{momp}), the current $\bm{J}_\perp$ is
\be
\bm{J}_\perp=-n_ee\bm{v}_\perp=-\frac{n_ee^2\bm{A}}{\gamma m_ec}.
\ee
Therefore, the electromagnetic wave equation could be written as
\be
\nabla^2\bm{A}-\frac{1}{c^2}\frac{\partial^2\bm{A}}{\partial t^2}=\frac{4\pi n_ee^2}{\gamma m_ec^2}\bm{A}.
\ee
Assuming that the wave satisfies $\bm{A}\propto\sin(\omega t-kx)$ and defining $\omega_p^2=4\pi e^2n_e/m_e$, one finally obtains \citep[e.g.][]{gib05,mac13}
\be
\omega^2=k^2c^2+\frac{\omega_p^2}{\gamma}.
\ee
This is the dispersion relation of an strong wave propagating in a plasma, which depends on the electron Lorentz factor $\gamma$ in details. 

Different from free electrons that have relativistic drift velocity, in plasma the space-charge potential was important in preventing the drift of electrons in the direction of the incident electromagnetic wave \citep{wal78}.
For an electron moving under a circularly polarized wave, 
the electron Lorentz factor is constant, i.e. \citep[e.g.][]{mac13}
\be
\gamma=\sqrt{1+\frac{a^2}{2}}.
\ee 
Therefore, for the strong circularly polarized waves, the dispersion relation is similar to the classical one with $\omega_p\rightarrow\omega_p/\sqrt{\gamma}$, where $\gamma=(1+a^2/2)^{1/2}$.

\end{document}